# Understanding the kinetics of static recrystallization in Mg–Zn–Ca alloys using an integrated PRISMS simulation framework

David Montiel[a], Philip Staublin[a], Supriyo Chakraborty[a], Tracy Berman[a], Chaitali Patil[a], Michael Pilipchuk[b], Veera Sundararaghavan[b], John Allison[a], Katsuyo Thornton[a,c]

[a]Department of Materials Science and Engineering, University of Michigan, Ann Arbor, MI 48109, United States

[b]Department of Aerospace Engineering, University of Michigan, Ann Arbor, MI 48109, United States

[c]Department of Nuclear Engineering and Radiological Sciences, University of Michigan, Ann Arbor, MI 48109, United States

## Abstract

Recrystallization is a phenomenon in which a plastically deformed polycrystalline microstructure with a high dislocation density transforms into another that has low dislocation density. This evolution is driven by the stored energy in dislocations, rather than grain growth driven by grain boundary energy alone. One difficulty in quantitative modeling of recrystallization is the uncertainty in material parameters, which can be addressed by integration of experimental data into simulations. In this work, we compare simulated static recrystallization dynamics of a Mg–3Zn–0.1Ca wt.% alloy to experiments involving thermomechanical processing followed by measurements of the recrystallization fraction over time. The simulations are performed by combining PRISMS software for crystal plasticity and phase-field models (PRISMS-Plasticity and PRISMS-PF, respectively) in an integrated computational materials engineering framework. At 20% strain and annealing at 350 °C, the model accurately describes recrystallization dynamics up to a mobility-dependent time scale factor. While the average grain boundary mobility and the fraction of plastic work converted into stored energy are not precisely known, by fitting simulations to experimental data, we show that the average grain boundary mobility can be determined if the fraction of plastic work converted to stored energy is known, or vice versa. For low annealing temperatures, we observe a discrepancy between the model and experiments in the late stages of recrystallization, where a slowdown in recrystallization kinetics occurs in the experiments. We discuss possible sources of this slowdown and propose additional physical mechanisms that need to be accounted for in the model to improve its predictions.

# 1. Introduction

Recrystallization significantly alters the microstructure and properties of metals and alloys by reducing dislocation density and restoring ductility after plastic deformation. In deformed polycrystals, new dislocation-free grains form and grow at the expense of grains with highly strained regions. In contrast to capillarity-driven grain growth, where the driving force for grain boundary (GB) migration is theorized to be proportional to mean curvature, in recrystallization, the driving force is proportional to the difference in stored energy due to dislocations between parent and recrystallized grains. Thus, as the dislocation-free grains grow, the total stored energy of the system decreases.

For magnesium, rolling often results in strong basal textures and highly anisotropic mechanical properties [1, 2]. Recrystallization is one of the processes that can improve poor formability after rolling by weakening the basal texture and reducing anisotropy in the mechanical properties [3]. In some Mg alloys, notably those including rare earth elements, it has been observed that recrystallization after rolling can generate weaker textures [4-11]. Promisingly, this behavior has also been observed in lower-cost alloys, such as those containing Zn and Ca [12-17]. Recent experiments on Mg–Zn–Ca alloys show that both the recrystallization kinetics and texture evolution have a strong dependence on alloy composition [16, 18]. However, the underlying mechanisms that give rise to this dependence are not well understood.

Given the significant effort involved in conducting recrystallization experiments and characterizing microstructures, the use of accurate models is key to understanding the mechanisms that govern recrystallization kinetics. The simplest approach to describe recrystallization kinetics is the Johnson-Mehl-Avrami-Kolmogorov (JMAK) model, which describes the dynamics of systems containing uniformly distributed nuclei or seeds that grow isotropically with a time-independent rate [19, 20]. The well-known Avrami equation relates the volume fraction of recrystallized grains, $X$, to the annealing time, $t$:

$$X = 1 - \exp\left(-\beta\left(\frac{t}{t_{1/2}}\right)^n\right), \qquad (1)$$

where $\beta = \ln(1/2)$, $t_{1/2}$ is the time for the recrystallized fraction to reach $0.5$, and $n$ is an exponent that depends on the dimensionality of the system and whether the nuclei/seeds form prior to or concurrently during growth. Assuming the former (the so-called site-saturation condition) in three dimensions, the constant $t_{1/2}$ can be related to the GB mobility, $M$, driving force, $\Delta G$, and number density of seeds, $N_0$ [21]:

$$t_{1/2} = \left(\frac{3 \ln 2}{4\pi N_0 (M\Delta G)^3}\right)^{1/3}. \qquad (2)$$

One of the more accurate and widely used methods to simulate microstructure evolution on the mesoscale is the phase-field (PF) method [22-27]. In PF models of recrystallization, grains are represented by order parameters that evolve in time to capture the motion of each GB. Several PF model approaches have been developed to simulate capillarity-driven grain growth [28-32], stored-energy-driven growth [33, 34], and static recrystallization [35-37]. However, achieving quantitative simulation predictions requires (1) that the model captures the physical mechanisms that drive the microstructure evolution and (2) that the physical parameters relevant to the mechanisms of interested are accurate.

One of the most challenging parameters to determine is the GB mobility. Given the crystalline nature of grains, both GB mobility and GB energy are, in general, anisotropic and depend on five macroscopic degrees of freedom defined by GB misorientation and GB plane inclination [38-41]. This poses a challenge in fully determining the mobility because of the large parameter space that needs to be explored even for one alloy, not to mention multiple alloy systems with a range of compositions. Atomistic simulations have given insight into the dependence of mobility on misorientation [42-44]. However, these findings have not yet been experimentally validated. In a key study, Zhang et al. [45] employed time-resolved X-ray diffraction contrast tomography measurements to experimentally determine the time-dependent morphology of grains within a polycrystalline Fe sample. Under the assumption of capillarity, i.e., that GB migration velocity is proportional to the local GB curvature, they fitted PF simulations of grain growth to determine the boundary mobility. However, they found no correlation between the GB mobility and the five macroscopic GB degrees of freedom and concluded that GB motion in polycrystals is governed by other factors. Moreover, recent studies show that shear-coupled GB migration plays a key role in GB motion [46] and that measuring GB migration in bicrystals [41, 47] does not necessarily represent the behavior observed in polycrystals.

For alloys, it is known that, even at small concentrations, solute elements can drastically affect GB mobility due to segregation-induced solute drag [38, 48, 49]. Lücke and Detert [50] first derived an expression for a steady-state GB migration velocity for systems with a high concentration of impurities. This velocity depends on the magnitude of the driving force, the impurity concentration at the GB, the solute diffusivity in the bulk, and temperature. Cahn later generalized this model for different GB driving force regimes [51]. In principle, Cahn's model can be applied to estimate GB mobilities in alloys. However, this estimation involves calculating a drag coefficient dependent on finite-temperature atomistic properties that are challenging to measure [52, 53].

The dynamics of recrystallization are largely determined by stored energy when the contribution from stored energy is much greater than the energy of the GBs. The GB migration velocity, $v$, can be expressed as [54]

$$v = M(F_s + F_\sigma),  \qquad (3)$$

where $F_s$ is the driving force due to the difference in stored energy arising from dislocations between parent and recrystallized grains and $F_\sigma$ is the driving force due to GB interfacial energy. When large plastic deformation introduces high stored energy, the contribution to the driving force from GB energy can be assumed to be negligible compared to that from the driving force from stored energy, such that $v \approx MF_s$. This allows more reliable estimation of GB mobilities when $F_s$ is known because, as evidenced by recent studies [41, 45-47], the dependence on curvature is less understood. The average GB mobility for different metals and alloys has been determined experimentally [49, 55, 56] by relating estimated average GB migration velocities and average driving forces from mean stored-energy values. However, in these studies, the effect of local variations in driving force is not captured due to averaging. In addition, the stored energy itself is difficult to determine from crystal plasticity finite element (CPFE) simulations due to uncertainty in the amount of plastic work converted to stored energy. Fractions of plastic work converted to stored energy range from 1-15%, depending on alloy composition and deformation conditions [57-59].

In this work, static recrystallization kinetics of a ZX30 Mg alloy (Mg–3Zn–0.1Ca wt.%) are experimentally characterized via electron backscatter diffraction (EBSD) measurements. Then, we combine CPFE and PF modeling to simulate static recrystallization dynamics for different initial deformations and annealing temperatures corresponding to the experiments. From the simulations, we obtain the recrystallized volume fraction vs. time and show that, for high temperature annealing (i.e., at 350 ºC), the simulated curves fit the experimental results up to a time scaling factor proportional to the average GB mobility. Because greater stored energy accelerates recrystallization, the time scaling factor also depends on the fraction of plastic work converted to stored energy. We show that if this fraction is known, the average GB mobility can be determined from the scaling factor. This paper is organized as follows. In Section 2, we describe the methods employed in this study. First, we describe the experimental procedures for thermomechanical deformation and annealing of the sample, as well as the steps involved in characterization using EBSD. Second, we focus on the CPFE method employed to simulate deformation and the postprocessing of the CPFE results to obtain the spatial distribution of stored energy after deformation. Third, we describe the PF model employed to simulate recrystallization during annealing, along with the method used to introduce seeds of recrystallized grains. In Section 3, we present the results of the recrystallization experiments and simulations, including a quantitative comparison of the recrystallization kinetics. In Section 4, we discuss how the average GB mobility can be obtained if the fraction of plastic work converted to stored energy is known. We also discuss possible extensions to the model, which could improve the quantitative fit to experiments at low annealing temperatures. Finally, in Section 5, we present our conclusions.

## 2. Methods

### 2.1 Experimental

Extrusions of ZX30 (Mg–3Zn–0.1Ca wt.%) were provided by CanMET Materials. The extrusions were annealed at 325 ºC for one hour before electrical discharge machining was used to extract specimens with dimensions of 20 mm × 15 mm × 10 mm. The shortest dimension was parallel to the extrusion direction. Figs. 1(a), (b), and (c) show the inverse pole figure orientation map, grain orientation spread (GOS) map, and crystallographic texture of the initial material after annealing and machining, respectively. These maps and pole figures show a fully recrystallized microstructure with weak basal texture. The number-average grain diameter was determined to be 20 µm based on image analysis of the EBSD micrograph in Fig. 1.

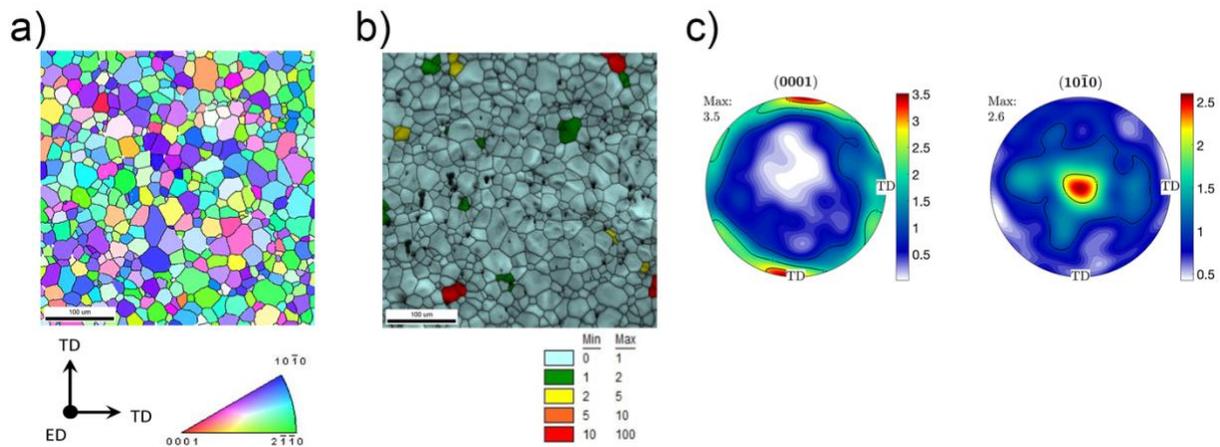

**Fig. 1.** Microstructure of the initial ZX30 extrusions as measured using EBSD: (a) inverse pole figure orientation map, (b) grain orientation spread map, and (c) basal and prismatic pole figures.

In order to examine the dynamics of recrystallization, we apply different thermomechanical treatment conditions to a series of specimens and characterize the microstructure. Thermomechanical deformation and most annealing treatments were performed in a Gleeble® Model 3500 thermo-mechanical simulator (Dynamic Systems Inc., Poestenkill, NY, USA). A K-type thermocouple was spot welded to the specimen. Plane strain compression was performed along the extrusion direction to the desired true strain (5%, 10%, or 20%). The true strain was calculated in the QuikSim™ software using a calibrated strain gauge within the Gleeble. The samples were deformed at 200 ºC with a strain rate of 0.5 s$^{-1}$. For samples annealed for less than 20 minutes, the specimen temperature was ramped up immediately after deformation from 200 ºC to the target annealing temperature at a rate of 5 ºC per second. The specimen was held at the annealing temperature for the target time and then quenched with forced air. For longer annealing treatments, the specimen was forced air quenched immediately following deformation and later annealed in a Thermolyne™ benchtop muffle furnace. In all cases, the annealing time was calculated as the total time above the 200 ºC deformation temperature. Thermal profiles of specimens annealed at 350 ºC are shown in Fig. 2.

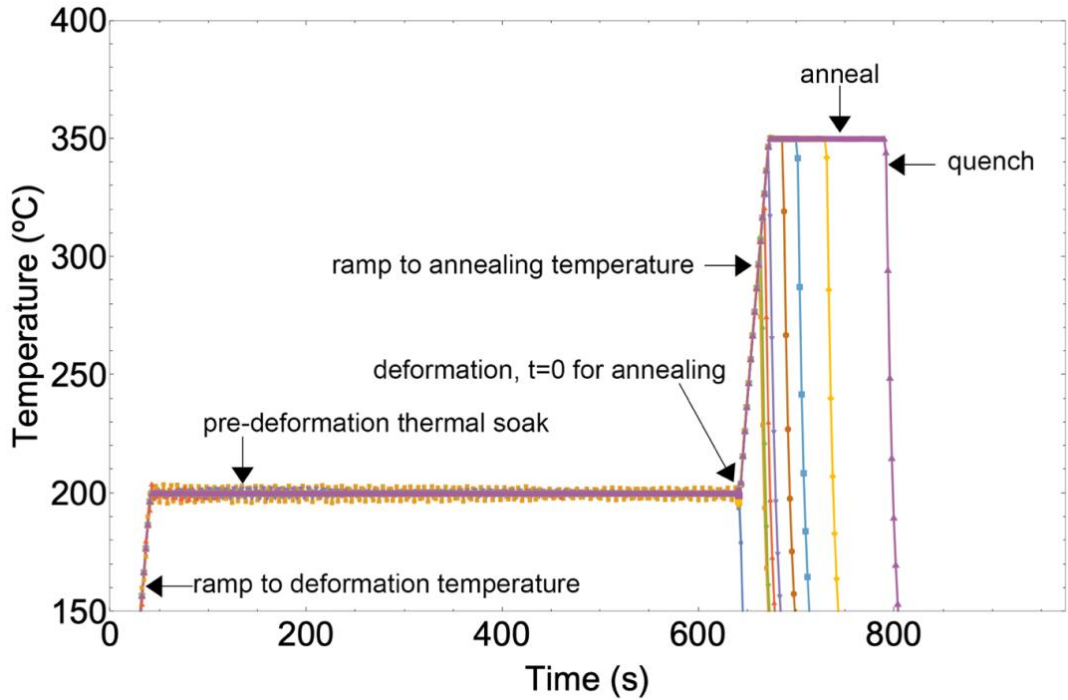

**Fig. 2.** Thermocouple profiles for multiple Gleeble specimens with a target annealing temperature of 350 ºC. The annealing duration is calculated as the time between deformation and the start of the air quench.

The Gleeble specimens were mounted in epoxy and were ground to mid-thickness along the compression direction. Polishing was conducted using successively finer diamond pastes with an oil-based lubricant. For some specimens, an additional final polishing step using fumed colloidal silica was utilized. The polished specimens were etched for approximately 5 s in a solution of 60 mL ethanol, 20 mL water, 15 mL glacial acetic acid, and 5 mL of nitric acid.

The amount of recrystallization that occurred during annealing was determined using EBSD GOS maps. EBSD analysis was conducted on a Tescan Mira3 scanning electron microscope, equipped with an EDAX Hikari EBSD detector, operating at 30 keV with a beam intensity of 20. All EBSD maps cover a scan region of 400 μm × 400 μm using a 1 μm step size. Data collection and data analysis were performed using the EDAX APEX and OIM Analysis software packages, respectively. Grains were identified based on a 5° misorientation tolerance. Points with a confidence index below 0.1 were excluded and are represented as black pixels in the resulting maps. For the purposes of this work, grains with a GOS less than 1° are considered recrystallized, while grains with a higher GOS are considered deformed [60]. A representative GOS map for the samples before deformation in the Gleeble is shown in Fig. 1(b), overlaid on grey-scale maps of the EBSD image quality (provided by the APEX Software); lighter regions have higher quality patterns, while darker regions have lower quality patterns that suggest more lattice distortion [61]. The number density, $N_v$, defined as the average number of grains per volume, was also estimated when the samples were nearly fully recrystallized using the stereological method by Saltykov described in Refs. [62-64] (see Supplementary Information).

## 2.2 Modeling

### 2.2.1 Crystal plasticity finite element method

The open-source PRISMS-Plasticity framework [65-67] was used for CPFEM simulations of thermomechanical deformation. PRISMS-Plasticity is a computationally efficient finite element solver for simulating a range of deformation conditions with applications such as fatigue [68, 69], twinning and detwinning [70], and texture evolution [71]. The strain-rate-dependent constitutive formulation used in this study is described below.

The deformation gradient tensor, $\mathbf{F}$, can be written as a multiplicative decomposition into the elastic and plastic deformation gradients, $\mathbf{F}^e$ and $\mathbf{F}^p$ respectively [72]:

$$\mathbf{F} = \mathbf{F}^e \mathbf{F}^p. \qquad (4)$$

Further, $\mathbf{L} = \dot{\mathbf{F}} \mathbf{F}^{-1}$ represents the velocity gradient in the deformed configuration, where $\dot{\mathbf{F}}$ is the time derivative of $\mathbf{F}$. Using Eq. (4), $\mathbf{L}$ can be written as

$$\mathbf{L} = \mathbf{L}^e + \mathbf{L}^p, \qquad (5)$$

where $\mathbf{L}^e = \dot{\mathbf{F}}^e \mathbf{F}^{e-1}$ and $\mathbf{L}^p = \mathbf{F}^e \dot{\mathbf{F}}^p \mathbf{F}^{p-1} \mathbf{F}^{e-1}$ are the elastic and plastic components of the velocity gradient in the deformed configuration, respectively [72]. The term $\dot{\mathbf{F}}^p \mathbf{F}^{p-1}$ represents the plastic velocity gradient in the intermediate configuration, $l^p$, which can also be expressed in terms of shear accumulation due to the activation of multiple slip systems. In this work, the influence of twin deformation is taken into account by considering twinning as a pseudo-slip activation. Therefore, $l^p$ is defined as

$$l^p = \dot{\mathbf{F}}^p \mathbf{F}^{p-1} = \sum_{\alpha=1}^{N_s+N_t} \dot{\gamma}^\alpha \mathbf{S}^\alpha, \qquad (6)$$

$$\mathbf{S}^\alpha = \mathbf{m}^\alpha \otimes \mathbf{n}^\alpha. \qquad (7)$$

Here, $\dot{\gamma}^\alpha$ is the shear rate on the $\alpha$-th slip system and $\mathbf{S}^\alpha$ is the Schmid tensor in the intermediate configuration. $N_s$ and $N_t$ represent, respectively, the number of slip systems and the number of twin systems. The symbols $\mathbf{m}^\alpha$ and $\mathbf{n}^\alpha$ represent, respectively, the unit vectors in the slip direction and slip plane normal direction in the intermediate configuration for slip systems and the twin plane normal and twin shear direction for twin systems.

The rate of shear accumulation on the $\alpha$-th slip system, $\dot{\gamma}^\alpha$ in Eq. (6), can be expressed using a power law formulation as [73]

$$\dot{\gamma}^\alpha = \dot{\gamma}_0 \left|\frac{\tau^\alpha}{s^\alpha}\right|^m \text{sgn}(\tau^\alpha), \tag{8}$$

where $\dot{\gamma}_0$ is the reference shear rate and $m$ is the stress exponent. The resolved shear stress, $\tau^\alpha$, can be obtained as

$$\tau^\alpha = \boldsymbol{\sigma} : \boldsymbol{S}^\alpha, \tag{9}$$

where $\boldsymbol{\sigma}$ is the Cauchy stress tensor and the colon denotes the tensor inner product (i.e., a double contraction). Finally, $s^\alpha$ represents the slip resistance on the $\alpha$-th slip system. The evolution of slip resistance can be captured from the hardening model,

$$\dot{s}^\alpha = \sum_{\beta=1}^{N_s} h^{\alpha\beta} |\dot{\gamma}^\beta|, \tag{10}$$

where the hardening matrix, $h$, defines the influence of the slip activity on the $\beta$-th slip system on the slip resistance of the $\alpha$-th slip system. The hardening moduli, $h^{\alpha\beta}$, are obtained using a power-law relationship:

$$h^{\alpha\beta} = \begin{cases} h_0^\beta \left[1 - \dfrac{s^\beta}{s_s^\beta}\right]^{a^\beta} & \text{if } \alpha = \beta \text{ (coplanar systems)} \\[1em] h_0^\beta q \left[1 - \dfrac{s^\beta}{s_s^\beta}\right]^{a^\beta} & \text{if } \alpha \neq \beta \text{ (non-coplanar systems)} \end{cases}, \tag{11}$$

where $h_0^\beta$ is the hardening parameter for the $\beta$-th slip system, $q$ is the latent hardening ratio, $s_s^\beta$ is the slip resistance at hardening saturation for the $\beta$-th slip system, and $a^\beta$ is an exponent that characterizes the sensitivity of the hardening moduli to the slip resistance of the $\beta$-th slip system.

Finally, the work done by the plastic deformation, $W^p$, is given by [74, 75]

$$W^p = \sum_{\alpha=1}^{N_s} \tau^\alpha \gamma^\alpha. \tag{12}$$

Most of the plastic work is dissipated into heat, while some fraction (in the range of ~1-15%) is converted into stored energy [57-59]. This fraction is challenging to determine because it depends on several factors, including alloy composition, temperature, applied strain, and strain rate. Therefore, we performed a sensitivity analysis using a range of fractions of plastic work converted to stored energy, from 0.75% to 15%.

The stress-strain response observed during the Gleeble® plane strain compression experiments was used to calibrate the crystal plasticity parameters. The plane strain boundary condition was applied using the velocity gradient tensor,

$$L_{Gleeble} = \begin{bmatrix} 0 & 0 & 0 \\ 0 & 0.5 & 0 \\ 0 & 0 & -0.5 \end{bmatrix} s^{-1}, \quad (13)$$

such that the strain rate for the Gleeble deformation and the plasticity simulations were consistent. Elastic constants of pure Mg as reported in Ref. [76] were used in this work. Basal $\langle a \rangle \{0001\}\langle 11\bar{2}0\rangle$, Prismatic $\langle a \rangle \{10\bar{1}0\}\langle 1120\rangle$, and Pyramidal $\langle a + c \rangle \{11\bar{2}2\}\langle 11\bar{2}3\rangle$ were considered as potential active slip systems. Tensile twin ($\{10\bar{1}2\}\langle\bar{1}011\rangle$) deformation was also considered as a pseudo-slip system. The crystal plasticity parameters, including slip resistance and hardening moduli, are listed in Tables 1 and 2.

**Table 1.** Elastic constants (in Voigt notation) and crystal plasticity parameters.

| Constant | Value | Unit | Reference |
|---|---|---|---|
| $C_{11}$ | 59400 | MPa | [76] |
| $C_{12}$ | 25610 | MPa | [76] |
| $C_{13}$ | 21440 | MPa | [76] |
| $C_{44}$ | 16400 | MPa | [76] |
| $\dot{\gamma}_0$ | 0.001 | s$^{-1}$ | Calibrated |
| $m$ | 0.1 | | Calibrated |
| $q$ | 1 | | Calibrated |
| $a^\beta$ | 1 | | Calibrated |

**Table 2.** Calibrated slip resistance and hardening moduli for the considered deformation modes.

| Deformation mode | $s_0^\beta$ (MPa) | $h_0^\beta$ | $s_s^\beta$ (MPa) |
|---|---|---|---|
| Basal | 2 | 200 | 30 |
| Prismatic | 25 | 300 | 400 |
| Pyramidal | 100 | 500 | 170 |
| Twinning | 60 | 700 | 500 |

## 2.2.2 Phase-field modeling

The phase-field (PF) model of Gentry and Thornton [34, 36] was used to simulate static recrystallization. The model is based on that of Moelans et al. [32] and employs multiple order parameters, $\eta_i$ ($i = 1, 2, \ldots N$), to represent different grains in the microstructure. For numerical efficiency, a single order parameter was allowed to represent multiple grains, ensuring that the grains of the same order parameter are sufficiently separated to limit artificial merging. The free energy of the polycrystalline domain is given by the functional,

$$F[\eta_1, \eta_2, \ldots, \eta_N] = \int_V [f_b + f_g + f_s] dV, \tag{14}$$

which contains the usual bulk and gradient free energy terms, $f_b$ and $f_g$, as well as an additional term, $f_s$, that accounts for the stored energy due to deformation. It is important to emphasize that $f_s$ represents a coarse-grained contribution from the energies of individual dislocations. The bulk and gradient terms are [32]

$$f_b = m_0 \left[ \frac{1}{4} + \sum_{i=1}^{N} \left( \frac{\eta_i^4}{4} - \frac{\eta_i^2}{2} \right) + \frac{3}{2} \sum_{i=1}^{N} \sum_{j<i}^{N} \eta_i^2 \eta_j^2 \right] \tag{15}$$

and

$$f_g = \frac{\kappa}{2} \sum_{i=1}^{N} |\nabla \eta_i|^2, \tag{16}$$

respectively. The gradient energy coefficient, $\kappa$, and the energy density coefficient, $m_0$, were assumed to be isotropic and independent of the order parameters; therefore, the GB energy is constant for all boundaries in the system in the absence of stored energy. This form of $f_b$ was chosen such that the equilibrium order parameter profile through a GB is symmetric in the absence of stored energy. The presence of nonuniform stored energy breaks this symmetry and modifies the order parameter profile through the GB, leading to an overestimation of the GB energy [36]. However, this effect was not significant in the present study (see Supplementary Information). The stored energy due to deformation was set based on the results of the crystal plasticity model in Section 2.2.1. As in previous works [34], the stored energy term is interpolated from the stored energy of each grain, $f_{s,i}$:

$$f_s = \frac{\sum_{i=1}^{N} \eta_i^2 f_{s,i}}{\sum_{j=1}^{N} \eta_j^2}. \tag{17}$$

The order parameters evolve in time according to a set of Allen-Cahn equations,

$$\frac{d\eta_i}{dt} = -L\frac{\delta F}{\delta \eta_i}, \tag{18}$$

where $L$ is the mobility coefficient for the evolution of the order parameter. Taking the variational derivative of the free energy functional, Eq. (14), the evolution equations can be expressed as

$$\frac{d\eta_i}{dt} = -L\left[m_0\left(\eta_i^3 - \eta_i + 3\eta_i \sum_{j\neq i}^{N} \eta_j^2\right) - \kappa\nabla^2\eta_i + 2\frac{\eta_i}{\sum_{j=1}^{N}\eta_j^2}(f_{s,i} - f_s)\right]. \tag{19}$$

A matrix-free finite element method was used to solve the weak form of Eq. (19), implemented in the open-source PRISMS-PF framework [77]. The equations were nondimensionalized following the same procedure as Huang et al. [34]. The energy was scaled by $m_0 = 1.33$ MJ / m$^3$, and length was scaled by $l_0$, which was chosen such that $l_0 = 2\Delta x_{min} = 1.5 \mu m$, where $\Delta x_{min}$ is the minimum grid spacing. Dividing energies and lengths by $m_0$ and $l_0$, respectively, results in nondimensional quantities, which were used in the simulations (Table 3). Dimensionless quantities are denoted with a superscripted star ($^*$). Time was scaled by the quantity $t_0 = 1/m_0 L$. The stored energy was scaled by $m_0$, resulting in the following expression for the nondimensional stored energy in terms of the dislocation density,

$$f_s^* = \frac{f_s}{m_0} = \frac{1}{2}Gb^2\rho\left(\frac{l_o}{4\sigma_{gb}}\right), \tag{20}$$

where $m_0 = 6\sigma_{gb}/l_{gb}$, $\sigma_{gb}$ is the GB energy, and $l_{gb}$ is the diffuse GB width. In recrystallized grains, dislocation density was assumed to be zero, i.e., $f_i^* = 0$ if $i$ is the index of a recrystallized grain. In the model, this was implemented by assigning $f_i = 0$ for all $i$ corresponding to recrystallized grains. The difference in stored energy drives the growth of the recrystallized grains at the expense of the deformed grains, reducing the total stored energy of the system. The fraction of recrystallized grains was calculated by integrating the order parameters representing recrystallized grains over the volume of the simulation domain.

All simulations employed the same representative microstructure as the initial condition constructed using DREAM.3D [78]. The experimentally measured grain size distribution in the as-deformed state was used when generating the artificial microstructure, resulting in a total of 463 grains in a $120^3$ μm$^3$ domain ($80^3$ voxels), which was used as input for the CPFEM calculations and the PF simulations. The PF simulations began with a resolution of $80^3$ voxels and used adaptive mesh refinement, resulting in locally higher resolution near the grain boundaries. The seed densities were set based on the experimental data, as explained in the next section. Four simulations were carried out under varying conditions: one with stored energy corresponding to 5% strain, one with stored energy corresponding to 10% strain, and two with stored energy corresponding to 20% strain. All four simulations have a different number density of recrystallized grains because the number density depends on both the applied strain and the annealing temperature. The two simulations with 20% strain were conducted with seed densities of recrystallized grains corresponding to annealing temperatures of 310 °C and 350 °C. Since the seed densities at 275 °C and 310 °C for 20% strain had similar values, the results for 275 °C and 310 °C used the same non-dimensional simulation. Because the composition-dependent GB mobility for Mg-Zn-Ca is not known, we assumed the temperature dependence appearing in the estimated temperature-dependent mobility of pure Mg [79]:

$$M \propto \exp\left(-\frac{69000}{RT}\right) \text{m}^4 / (\text{J} \cdot \text{s}), \tag{21}$$

where $R$ is the gas constant in J mol$^{-1}$ K$^{-1}$ and $T$ is the absolute temperature in K. This expression results in a ratio of mobility between 350 °C and 310 °C of 2.5. Similarly, the ratio of mobility between 350 °C and 275 °C is found to be 6.2. The simulation time scale, $t_0$, for 275 °C and 310 °C is determined by multiplying $t_0$ for 350 °C by these ratios. The mobility coefficient, $L$, was set based on the GB mobility: $L = (4M)/(3l_{gb})$, where $M \approx 8 \cdot 10^{-13}$ m$^4$ J$^{-1}$ s$^{-1}$ at 350 °C, which is approximately the value extrapolated from the fit given by Okrutny et al. [79].

Table 3. Phase field simulation parameters.

| Parameter | Value | Dimensionless Value |
|---|---|---|
| Grain boundary energy, $\sigma_{gb}$ | 0.5 J / m$^2$ | - |
| Grain boundary mobility, $M$ | $8 \cdot 10^{-13}$ m$^4$ / (J·s) | - |
| Mg shear modulus, $G$ | 17 GPa | - |
| Basal Burgers vector length, $b$ | 0.321 nm | - |
| Diffuse grain boundary width, $l_{gb}$ | 2.25 μm | $l_{gb}^* = \dfrac{l_{gb}}{l_0} = 1.5$ |
| Length scale, $l_0$ | 1.5 μm | - |
| Energy density coefficient, $m_0$ | 1.33 MJ / m$^3$ | $m_0^* = \dfrac{m_0}{m_0} = 1$ |
| Gradient energy coefficient, $\kappa$ | $8.44 \cdot 10^{-7}$ J / m | $\kappa^* = \dfrac{\kappa}{m_0 l_0^2} = 0.28125$ |
| Mobility coefficient, $L$ | $4.74 \cdot 10^{-7}$ m$^3$ / (J·s) | $L^* = \dfrac{L}{L} = 1$ |
| Stored energy, $f_s$ | - | $f_s^* = \dfrac{f_s}{m_0}$ |
| Position, $x$ | - | $x^* = \dfrac{x}{l_0}$ |
| Time, $t$ | - | $t^* = \dfrac{t}{t_0} = m_0 L t$ |

### 2.2.3 Method for seeding recrystallized grains

In static recrystallization, recrystallized grain seeds can be assumed to be already present at the start of recrystallization, and thus we generated the seeded microstructure as an initial condition. In this work, the number density of seeds for the simulations, $N_v^s$, was chosen to match the estimated number density of recrystallized grains from experiments at the point of full recrystallization, $N_v$. This is based on the assumption that seeds that vanish due to coalescence in the early stages of recrystallization are unlikely to contribute to the overall recrystallization dynamics at later stages. In previous experiments, pristine grains are more frequently observed to form at GB sites with high stored energy [48, 80]. Thus, a local probability density was used such that the most likely seeding sites to correspond to regions of high probability of finding a subgrain of supercritical size [81, 82],

$$P_{\text{sub}} = \int_{r_c}^{\infty} p(r)\,dr = \exp\left(-\frac{\pi r_c^2}{4\bar{r}^2}\right), \tag{22}$$

where $\bar{r}$ is the mean subgrain size calculated as

$$\bar{r} = \frac{C\sqrt{Gb^2}}{\sqrt{2\,f_s}}, \tag{23}$$

and $r_c$ is the critical radius for a subgrain to become a recrystallized seed,

$$r_c = \frac{2\sigma_{gb}}{f_s}. \tag{24}$$

In Eq. (23), $C$ is a dimensionless proportionality constant related to the entanglement of dislocations to form subgrain boundaries. In this work, this constant was adjusted such that the number density of critical subgrains reproduced the number density of recrystallized grains near full recrystallization, estimated from the experimental microstructures.

The placement of critical subgrains into the deformed microstructure was performed as follows: first, the deformed structure was evolved using Eq. (19) for $t^* = 0.5$ such that the phase fields attain proper diffuse-interface profiles. Next, the simulation was paused, and the elements belonging to the GB region were identified as those where $\sum_{i=1}^{N} \eta_i \eta_j > 0.15$, which corresponds to regions with order parameters having the range between $\approx 0.2$ and $\approx 0.8$. The probability was then computed in each GB element using Eq. (22), and if a uniform random number between 0 and 1 was below the calculated probability, the element was marked for seeding. Finally, elements were removed from the list of potential seeding sites such that no two seeds were within two radii from one another. Spherical subgrain seeds with a diameter of 6 μm were then placed, the mesh was refined, and the simulation was allowed to proceed, treating the time of seeds placement as $t^* = 0$. To examine the effect of random placement of the critical subgrains, each simulation was repeated twice using different random seeds as input to the pseudorandom number generator.

Both the CPFEM and PF simulations were performed on the Anvil supercomputer at the Rosen Center for Advanced Computing in Purdue University, through allocation MSS160003 from the Advanced Cyberinfrastructure Coordination Ecosystem: Services & Support (ACCESS) program [83]. Data used in this manuscript is made available through Materials Commons [84].

## 3. Results

## 3.1 Experimental characterization

GOS maps of the material strained to 20% and then annealed with a target temperature of 350 ºC are shown in Fig. 3. In the as-deformed condition, there are no grains with a GOS less than 1°, indicating that dynamic recrystallization (DRX) is not significant in the ZX30 alloy subjected to the thermomechanical processing conditions used in this work. ZX30 has also been shown to be resistant to DRX even at higher temperatures (e.g., 350 °C) [16, 85], where DRX is more likely to occur. Since the initial state consists of only deformed grains, it is a suitable starting point for monitoring the evolution of static recrystallization. Under the assumption that the microstructure in the GOS maps contains a representative area to capture the size and shape distribution of grains, the recrystallized volume fraction was taken as the area fraction of grains for which GOS is more than 1°. The as-deformed sample has a strong basal texture, as evidenced by the strong alignment of the basal plane normal direction with the loading direction, shown in Fig. 4(a).

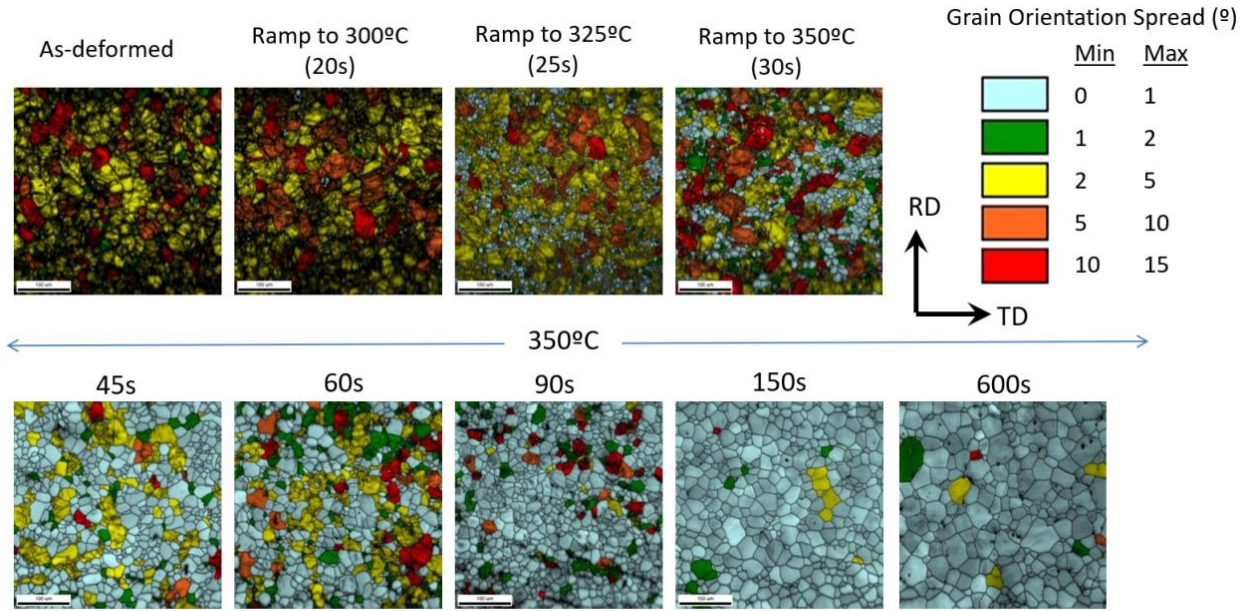

**Fig. 3.** GOS maps showing the grain size and extent of static recrystallization as a function of annealing time for specimens with a target annealing temperature of 350 ºC. The light blue grains are considered to be recrystallized. The scale bar is 100 μm and is the same for each image.

Some small, low GOS grains are observed at 300 ºC, but widespread recrystallization is not seen until 25 s after deformation, when the specimen temperature has increased to 325 ºC. By the time the sample is heated to the target annealing temperature of 350 ºC, approximately a quarter of the microstructure is already recrystallized. The microstructure is nearly fully recrystallized within 150 s. Fig. 4(b) shows the texture at 600 s annealing time, which corresponds to the last frame in Fig. 3. The texture after recrystallization is weak and remains so even during grain growth, suggesting a lack of preferential growth direction of the recrystallized grains.

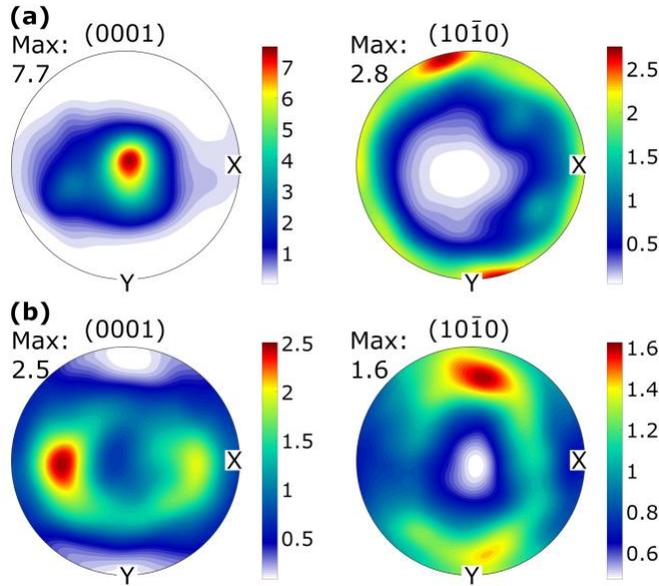

**Fig. 4.** Basal (left) and prismatic (right) pole figures showing the texture (a) of a sample in the as-deformed condition and (b) of the sample deformed to 20% true strain and annealed at 350 °C for 600 s. The compression direction is out of the page. The transverse direction is along the longest sample dimension and is labeled X, while Y is the rolling direction in the Gleeble plane strain compression experiment.

We now examine the effect of varying the applied strain (Fig. 5) and temperature (Fig. 6) on the rate of recrystallization. Plots of the recrystallized volume fraction vs. time above 200 ºC are shown in Fig. 5 for 5%, 10%, and 20% applied true strain for a fixed target annealing temperature of 350 ºC. The experimental data exhibits some scatter, since each data point represents a distinct sample and the initial microstructure varies slightly between samples; however, the time to reach nearly complete recrystallization is approximately doubled for 5% applied strain compared to 20% applied strain. The recrystallization volume fraction as a function of time above 200 ºC is shown in Fig. 6 for target annealing temperatures of 275 ºC, 325 ºC, and 350 ºC for a fixed applied true strain of 20%. The recrystallization rate increases with annealing temperatures because the average GB mobility increases with temperature. The annealing time required to reach full recrystallization increases by approximately an order of magnitude when annealing at 275 ºC compared to 350 ºC. The estimated number density of recrystallized grains near full recrystallization, $N_v$ (reported in Table 4), increased concomitantly with applied strain; a greater number density of recrystallized grains was also observed at 310 °C compared to 350 °C for 20% applied true strain.

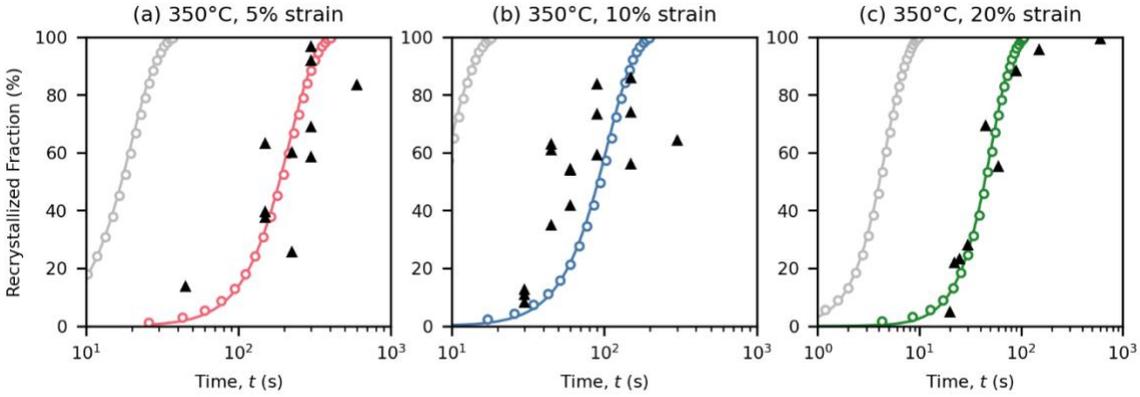

**Fig. 5.** Comparison of experimental and simulated recrystallization fractions vs. annealing time (black triangles vs. colored points, respectively) at an annealing temperature of 350 ºC for (a) 5% strain, (b) 10% strain, and (c) 20% strain. The colored lines are least-squares fits of the JMAK equation to the PF simulation data. The time scale is adjusted by a factor of 10.85 for each simulation such that the best fit between simulation and experiment is achieved for the case at 350 °C with 20% strain. The grey lines show the simulation data with the mobility of pure Mg.

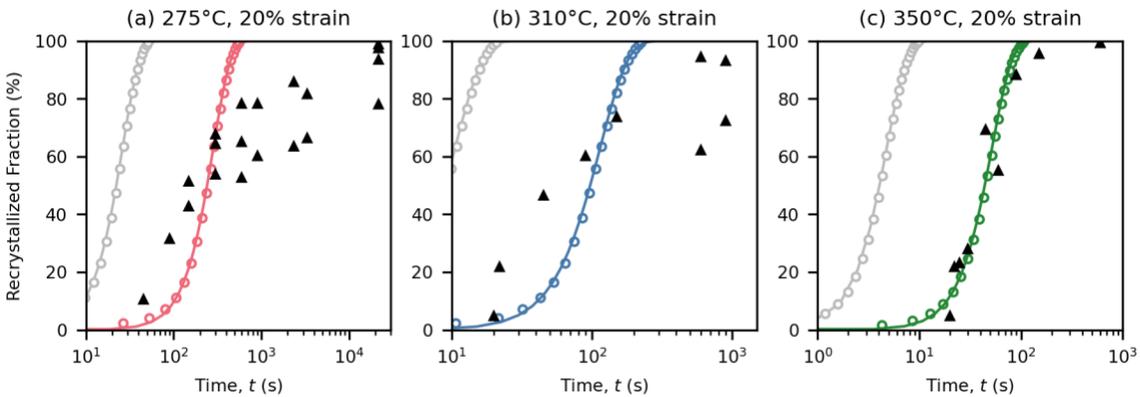

**Fig. 6.** Comparison of experimental and simulated recrystallization fractions vs. annealing time (black triangles vs. colored points, respectively) for 20% applied strain with annealing temperatures of (a) 275 ºC, (b) 310 ºC, and (c) 350 ºC. Plot (c) is identical to Fig. 5c. The colored lines are least-squares fits of the JMAK equation to the PF simulation data. The time scale is adjusted by a factor of 10.85 for each simulation such that the best fit between simulation and experiment is achieved. The grey lines show the simulation data with the mobility of pure Mg.

**Table 4.** Number density of grains determined from experiment at specific annealing times, calibrated seeding parameters, and number density of recrystallized seeds in the simulations.

| Condition | $N_v$ (μm⁻³) experiment | Selected time (s) | Seeding parameter, $C$ | $N_v^s$ (μm⁻³) simulation |
|---|---|---|---|---|
| 350 °C, 20% strain | $1.5 \cdot 10^{-4}$ | 150 | 3.7 | $1.5 \cdot 10^{-4}$ |
| 350 °C, 10% strain | $7.4 \cdot 10^{-5}$ | 90 | 5.45 | $7.8 \cdot 10^{-5}$ |
| 350 °C, 5% strain | $3.0 \cdot 10^{-5}$ | 600 | 8.1 | $3.3 \cdot 10^{-5}$ |
| 310 °C, 20% strain | $2.3 \cdot 10^{-4}$ | 600 | 4.3 | $2.1 \cdot 10^{-4}$ |
| 275 °C, 20% strain | $2.1 \cdot 10^{-4}$ | 21600 | - | - |

## 3.2 Integrated CPFE-PF simulations

The growth of recrystallized grains in the PF simulations is visualized by plotting the stored energy overlaid with the GBs (Figs. 7(a)-(c) for 310 °C, and Figs. 7(d)-(f) for 350 °C). Because the probability of subgrain seeding depends on the stored energy and the presence of GBs, critical subgrains are mostly placed in regions with high stored energy; in addition, the seeds occur most frequently on GBs and triple junctions. The subgrains grow rapidly into these regions, consuming the deformed microstructure. Curvature-driven grain growth occurs simultaneously with stored-energy-driven growth, but the stored energy contributes more to the driving force for GB migration. Regions with the largest stored energy tend to be consumed first, as expected, since the magnitude of the driving force for GB motion in the PF model is proportional to the local stored energy. This can be observed, for example, in regions highlighted by red circles in Figs. 7(a) and 7(b), but also in Figs. 7(d) and 7(e).

The average grain size in the fully recrystallized structures depends on the number density of critical subgrains, which varies with the annealing temperature (Table 4). The initial grain structure is the same in both simulations, except for the number of critical subgrains (Figs. 7(a) and 7(d)). Because the number density of recrystallized seeds is approximately 30% greater at 310 °C than 350 °C and most seeds survive to near full recrystallization, the average grain size is smaller after annealing at 310 °C compared to 350 °C (Figs. 7(c) and 7(f), respectively). As previously mentioned, grain growth among the recrystallized grains is slow compared to the stored-energy driven growth. For example, the shape of the triangular grain embedded among recrystallized grains maintains similar size and morphology (blue circle in Fig 7(e) and 7(f)). In addition, the time to reach full recrystallization depends on the average GB mobility, which varies with annealing temperature, as will be discussed in the next subsection. Similar recrystallization fractions are shown in Figs. 7(c) and 7(f), but the time to reach the depicted microstructures more than doubles for annealing at 310 °C compared to annealing at 350 °C.

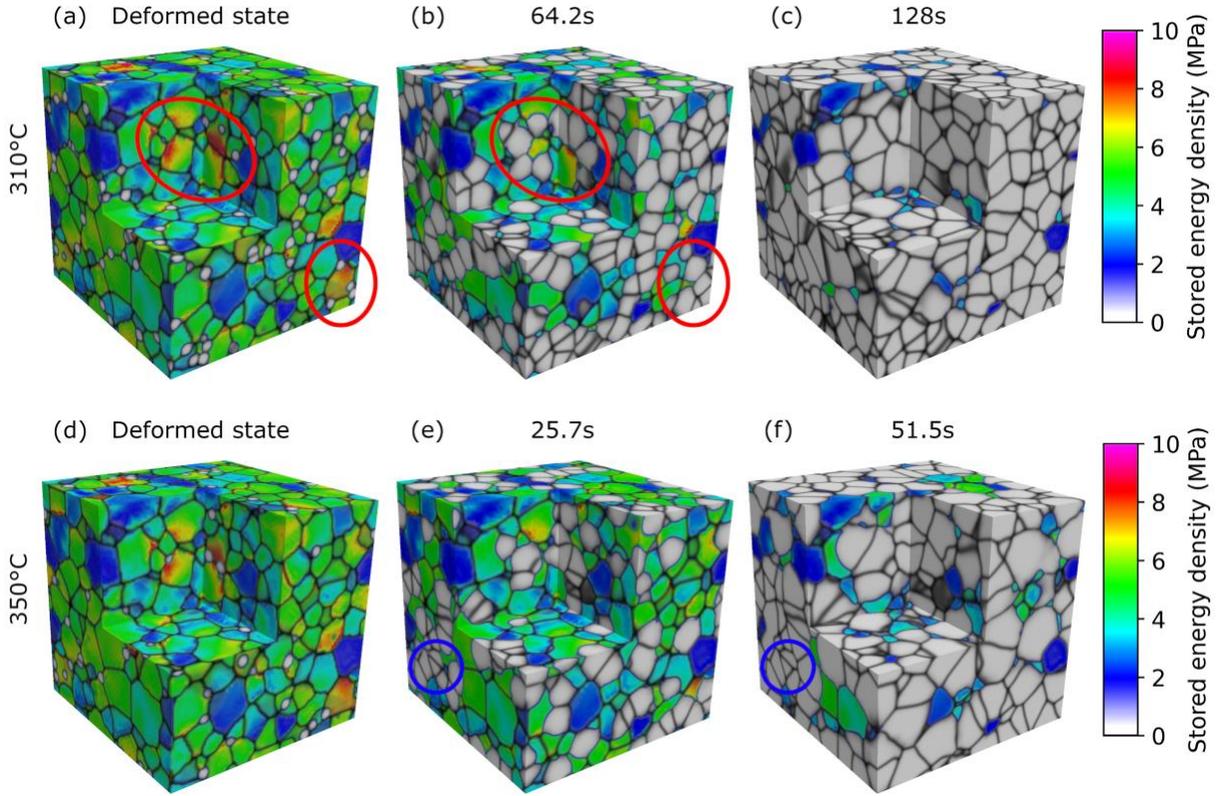

**Fig. 7.** Microstructure evolution during PF simulation of static recrystallization at (a)-(c) 310 °C and (d)-(f) 350 °C with 20% true strain; $t = 0\ s$ marks the seeding of critical subgrains in the deformed structure after a short relaxation of the initial condition. The recrystallized fractions are (b) 23%, (c) 70%, (e) 18%, and (f) 60%. Grain boundaries are highlighted in black by overlaying the quantity $1 - \sum_{i=1}^{N} \eta_i^3$. The timescale is set using the fitted mobility as described in Section 3.3. Values that would appear magenta based on the color scale are embedded within the volume and are therefore not visible.

### 3.3 Fitting of the grain boundary mobility

We perform a sensitivity analysis for the fraction of plastic work converted to stored energy to determine how the assumed conversion fraction affects the resulting GB mobility obtained by fitting simulation and experiment. We repeat the simulation using 20% applied strain and annealing at 350 °C with varying fractions of plastic work converted into stored energy, from 0.75% to 15%, and then obtain the simulation time scale for each case. To do so, the root mean squared error (RMSE) between the experimental data and the JMAK curve fitted to the simulation data is minimized with respect to the simulation time scale, $t_0$. The fitted simulation time scale is then used to compute the GB mobility.

Because the crystal plasticity constitutive model used in this study predicts the plastic work performed and not the dislocation density, a critical assumption is the amount of plastic work converted to the stored energy driving force for recrystallization. Consequently, different assumed plastic work conversion fractions result in different values for the fitted average GB mobility. However, these two quantities are related by the following asymptotic result of the PF model [36, 48], which relates the GB velocity to the mobility ($M$), surface energy ($\sigma_{gb}$), curvature ($K$), and stored energy driving force ($f_s$):

$$v \propto M(\sigma_{gb} K + F_s). \tag{25}$$

If $\sigma_{gb} K \ll F_s$, then $v \approx M F_s$ and the fitted mobility is therefore inversely related with the stored energy driving force for a given GB velocity. To test this hypothesis, we perform a sensitivity analysis, wherein we vary the fraction of plastic work converted to stored energy from 0.75% to 15% and repeat the simulation workflow. The effect of randomness in seeding recrystallized grains is examined by repeating each simulation with two different random-number-generator seeds. The stored energy conversion fraction demonstrates an inverse relationship with the resulting average GB mobility (Fig. 8), as expected based on Eq. (25). In addition, the spread in fitted mobility due to random seeding of recrystallized grains is small compared to the effect of the stored energy conversion fraction. Above 3% plastic work converted to stored energy, the stored energy term becomes the dominant term.

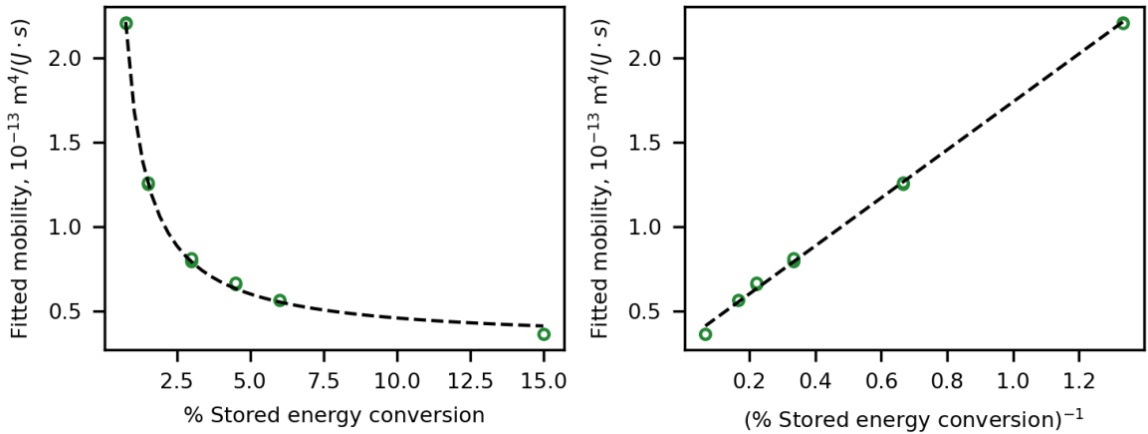

**Fig. 8.** Fitted average grain boundary mobility for ZX30 from PF simulations of static recrystallization at 350 °C and 20% applied strain, as a function of the amount of plastic work converted to stored energy. Both plots show the same data, but the right plot inverts the horizontal axis to show the linear inverse relationship. The black dashed line is the linear least-squares regression.

Using 3% conversion of plastic work to stored energy, the fitted average GB mobility for the ZX30 alloy is approximately 11 times that of pure Mg. To examine the effect of varying deformation temperature and applied strain, we employ the fitted mobility at 350 °C and 20% applied strain for simulations at other temperatures and applied strains. The mobility is adjusted with temperature according to Eq. (21) as described at the end of Section 2.2.2. The simulated recrystallization fractions versus time demonstrate varying agreement with experimental measurements of recrystallization as determined by GOS <1º (Figs. 5 and 6). More scatter in recrystallization fraction is observed between samples in the experiment at 5% and 10% applied strain (Fig. 5a and Fig. 5b, respectively). At 20% applied strain and 275 °C annealing temperature, the experimentally measured recrystallization rate deviates from the JMAK curve at longer annealing times (Fig. 6a). Therefore, the result at 20% applied strain and 350 °C annealing temperature was used to perform the fitting. Thus, this case is shown in both Fig. 5 and Fig. 6 (namely, Fig. 5c and Fig. 6c, respectively). The RMSE between the experimental data and the JMAK curve fit to the simulation data with 20% applied strain are 24%, 22%, and 10% for annealing at 275 °C, 310 °C, and 350 °C, respectively. For annealing at 350 °C with 5% and 10% applied strain, the RMSEs are 19% and 28%, respectively. The best quantitative agreement is obtained for annealing at 350 °C with 20% applied strain.

## 3.4 Effect of applied strain and annealing temperature

The effect of the amount of applied strain was investigated in the simulations by using different initial stored energy derived from CPFEM simulations at differing applied strain. The number density of seeds is adjusted for each simulation based on the number density of recrystallized grains in the corresponding experiments (Table 4). Lower applied strain results in a smaller number density of recrystallized grains due to a lower driving force for the formation of subgrain boundaries. To adjust the number density of seeds in the simulations, the constant parameter $C$ is changed in Eq. (23); the values used in the simulations are also given in Table 4.

With increasing strain, recrystallization both begins and reaches its maximum at earlier times. The fitted Avrami exponents are 2.49, 2.46, and 2.20 for the simulations at 5%, 10%, and 20% true strain, respectively. The time to reach 50% recrystallization fraction is 186 s, 91.1 s, and 44.0 s for the simulations at 5%, 10%, and 20% true strain, respectively (Table 5).

**Table 5.** Fitted Avrami parameters for recrystallization simulations under different conditions

| Condition | Avrami exponent, $n$ | Time to 50% recrystallization, $t_{1/2}$ (s) |
|---|---|---|
| 350 ºC, 20% strain | 2.20 | 44.0 |
| 350 ºC, 10% strain | 2.46 | 91.1 |
| 350 ºC, 5% strain | 2.49 | 186 |
| 310 ºC, 20% strain | 2.20 | 97.2 |
| 275 ºC, 20% strain | 2.20 | 241 |

At lower annealing temperatures, the experimentally observed rate of recrystallization decreases (Fig. 6). To simulate recrystallization at lower temperatures, one additional simulation was conducted at 20% applied strain with a greater number density of seeds, to reflect the increased number density of recrystallized grains in the experiment at 310 °C. Because no significant difference was observed in the number density of seeds between 310 °C and 275 °C, the same simulation was used to describe both temperatures, but rather only the time scale is adjusted between the 310 °C and 275 °C simulations to reflect the concomitant decrease of mobility with decreased temperature. Recrystallization is predicted to be faster with increasing temperature, which is expected due to the increase in GB mobility. However, the Avrami exponent remains 2.20 at both 310 °C and 350 °C (and the same at 275 °C), since the seeding condition is similar and only the time scale is shifted. The time to reach 50% recrystallized fraction is determined to be 241 s, 97.2 s, and 44.0 s for the simulations at 275 °C, 310 °C, and 350 °C, respectively. At 275 °C, the experimentally observed rate of recrystallization appears to decrease with time (Fig. 6a), indicating that either mobility or driving force (or both) is reducing with time. Such changes are not captured by the PF simulation, and thus this experimental observation is not reproduced (Fig. 6a).

## 4. Discussion

Calibration of the GB mobility in PF simulations against experimental measurements of recrystallization fraction in the ZX30 alloy yields quantitative agreement at 350 ºC with 20% applied strain (Figs. 5 and 6). For 5% and 10% applied strain, the agreement is only qualitative, and at lower annealing temperatures of 275 ºC and 310 ºC, the agreement worsens. The calculated average GB mobility is 4–15 times lower than that measured in pure Mg depending on the assumed amount of plastic work that is converted to stored energy. This suggests that the presence of alloying elements in this alloy reduces the mobility by about one order of magnitude. Solute drag can impede GB motion, explaining the slower kinetics of recrystallization in ZX30, as compared to pure Mg. Direct experimental measurement of the GB mobility is difficult; however, measurement of the static recrystallization rate by identifying recrystallized grains based on GOS is more straightforward. By fitting the PF simulation timescale to the experimental recrystallization measurements, the average GB mobility can be estimated, provided that the fraction of plastic work converted into stored energy, and thus the magnitude of the driving force due to stored energy, is known. This mobility can be used as a means to quantify the effect of solute on GB motion.

A key assumption in the computation of the dislocation density is the conversion fraction of plastic work into stored energy, since the amount of stored energy impacts the driving force for recrystallization. However, the amount of plastic work converted into stored energy is challenging to determine both experimentally and through computations. Nevertheless, even without knowing the exact conversion fraction, the range of possible average GB mobility values can be estimated with the present methodology. For example, as shown in Fig. 8, if only 1.5% of the plastic work was converted to stored energy, rather than the 3% baseline value we assumed, the driving force for growth of recrystallized grains would be reduced by approximately half. Since the driving force scales linearly with the conversion fraction, the fitted mobilities are inversely proportional to the assumed conversion fraction. However, since the conversion fraction may vary with applied strain [86], the GB mobility estimated for one condition may not generalize to other applied strains. Future work investigating the strain- and temperature-dependence of the conversion of plastic work into stored dislocations could allow for generalization of the estimated average GB mobility to other processing conditions. For example, the broadening of X-ray diffraction peaks can be used to measure the total dislocation density after plastic deformation, which can be separated into statistically stored and geometrically necessary dislocations when combined with electron backscatter diffraction [87].

The agreement of the simulations and experiments for the 350 ºC annealing temperature is achieved despite the simplifying assumptions in the PF model. For example, the interaction of individual dislocations with grain boundaries and with each other is coarse-grained into the stored energy field. In addition, the model does not account for variation in GB energy with misorientation or inclination. Although both GB energy and mobility are strongly anisotropic and dependent on misorientation, during recrystallization the stored energy contributes significantly more to the driving force for boundary motion compared to capillarity. Moreover, if anisotropy in the GB energy and mobility significantly affected the growth of recrystallized grains, a preferential growth direction would be expected, leading to anisotropic grain morphology and the development of stronger texture after recrystallization than observed in the present work (Figs. 3 and 4). Therefore, anisotropy is likely less important to the overall recrystallization dynamics. Nevertheless, accounting for anisotropic GB properties may potentially improve the predictions of the model. However, for each GB, these properties generally depend on the five degrees of freedom defined by the misorientation and inclination of GBs. To our knowledge, the precise dependence of GB energy and mobility on the five-dimensional space of GB crystallography has not been established even for pure Mg, much less for the ternary alloy considered in this work. Therefore, the inclusion of anisotropy-dependent terms within the model is especially challenging. It is also worth noting that the recrystallization dynamics in both simulations and experiments for different initial deformations and an annealing temperature of 350 ºC is accurately described by the JMAK equation. This behavior suggests that the underlying assumptions of the JMAK model are valid for recrystallization kinetics under the aforementioned conditions.

The current model also does not account for solute partitioning, second-phase particles, or static recovery; therefore, the current methodology cannot necessarily be generalized to new alloy systems without first considering the impact of these factors. For instance, depending on alloy composition, the precipitation of second-phase particles may either impede recrystallization by pinning grain boundaries or accelerate recrystallization by promoting nucleation of subgrains [48, 88]. Another factor to consider is the texture of the parent microstructure, which affects the orientation of the recrystallized grain seeds.

Introducing additional complexities into the model could improve accuracy, but at the cost of increased computational resources. In particular, at low temperature, the experimental data deviates from the simulation prediction towards longer time and larger recrystallization fraction. Because of the slower kinetics at 275 °C, solute drag may play a larger role due to a longer time for solute accumulation at grain boundaries. In addition, static recovery may contribute to a reduction in the stored energy over time as dislocations annihilate by climb. The PF model could be extended to include a recovery term based on rate laws for static recovery; however, the rate of recovery would need to be calibrated, for example, through dislocation dynamics simulations [89].

## 5. Conclusions

Comparison of the simulations with experimental data shows that, across various applied strains and at high annealing temperatures, the model reproduces the recrystallization dynamics accurately, with an appropriate time-scaling factor determined by the mobility and the fraction of converted plastic work. For each value of this conversion fraction, the average grain boundary mobility in a Mg–3Zn–0.1Ca (wt.%) alloy was estimated by fitting phase field simulations of static recrystallization to experimental measurements of the recrystallization fraction versus annealing time. The following conclusions are reached based on the experiments and simulations:

(1) No evidence of dynamic recrystallization was observed during electron backscatter diffraction of samples in the as-deformed condition, supporting the application of static recrystallization models to the ZX30 alloy.
(2) At larger applied strains, the increased stored energy provides a greater driving force for static recrystallization, leading to faster recrystallization kinetics. The driving force due to stored energy is significantly larger than that due to grain boundary curvature.

(3) The curves for recrystallization fraction vs. time calculated from phase-field simulations with the estimated average mobility reasonably agree with the experimental data for different values of the initial strain and an annealing temperature of 350 ºC. The discrepancy at other temperatures is likely due to solute drag that impedes grain boundary motion.

The calibration of simulation parameters using experimental data enables estimation of material properties that are difficult to measure directly, such as the grain boundary mobility. As mesoscale models of microstructure evolution become more sophisticated and incorporate additional physics, the present approach can be fully integrated into a workflow and can be applied to more complex engineering alloys, accelerating materials design and optimization.

## Acknowledgments

This work was supported by the U.S. Department of Energy Office of Basic Energy Sciences Division of Materials Science and Engineering under Award #DE-SC0008637 as part of the Center for PRedictive Integrated Structural Materials Science (PRISMS). This work used the Anvil supercomputer at the Rosen Center for Advanced Computing in Purdue University through allocation MSS160003 from the Advanced Cyberinfrastructure Coordination Ecosystem: Services & Support (ACCESS) program, which is supported by U.S. National Science Foundation grants #2138259, #2138286, #2138307, #2137603, and #2138296.

Supplementary Information for Understanding the kinetics of static recrystallization in Mg-Zn-Ca alloys using an integrated PRISMS simulation framework

D. Montiel, P. Staublin, S. Chakraborty, T. Berman, C. Patil, M. Pilipchuk, V. Sundararaghavan, J. Allison, and K. Thornton

18 February 2026

*S1. Sensitivity Analysis for Stored Energy Conversion*

Because the crystal plasticity constitutive model used in this study predicts the plastic work performed and not the dislocation density, a critical assumption is the amount of plastic work converted into the stored energy driving force for recrystallization. The chosen ratio of plastic work converted to stored energy affects the fitted grain boundary mobility and is therefore a confounding variable in this study. However, the two variables are related by the asymptotic result of the phase field model [1,2], which relates the grain boundary velocity to the mobility ($M$), surface energy ($\gamma$), curvature ($K$), and stored energy driving force ($f_{stored}$):

$$v \propto M(\gamma K + f_{stored}). \quad (S1)$$

If $\gamma K \ll f_{stored}$, then $v \approx M f_{stored}$ and the fitted mobility is therefore inversely related with the stored energy driving force for a given grain boundary velocity. To test this hypothesis, we perform a sensitivity analysis, wherein we vary the fraction of plastic work converted to stored energy and repeat the simulation workflow. The effect of randomness in seeding recrystallized grains is controlled by repeating each simulation with two different pseudorandom number generator seeds. Because the literature reports a wide range of possible ratios of plastic work to stored energy (1-15%) [3-5], we repeat the study for 0.75%, 1.5%, 3%, 4.5%, 6%, and 15% of plastic work converted to stored energy. The grain boundary mobility obtained as a result of fitting to experimental data demonstrates an inverse relationship with the assumed fraction of plastic work converted to stored energy (Fig. 8 in the main article), as expected based on Eq. (S1) under the assumption $\gamma K \ll f_{stored}$. In addition, the spread in fitted mobility due to random seeding of recrystallized grains is small compared to the effect of the stored energy conversion ratio.

Finally, we examine the assumption that $\gamma K \ll f_{stored}$ in two ways. First, the linear fit in Fig. 8 (main article) is repeated, excluding the points at 0.75% and 1.5% stored energy conversion. These conversion fractions correspond to the smallest $f_{stored}$ so these points are most likely to approach the regime where $\gamma K \approx f_{stored}$. This new fit exhibits an improved correlation coefficient compared to that using all the data ($R^2$ = 0.926 vs. $R^2$ = 0.988). In addition, the excluded data at 0.75% and 1.5% stored energy is below the new linear fit, suggesting an increasing importance of the curvature contribution to grain boundary motion as shown in Fig. S1. Next, we compare the contribution from the driving force due to curvature to the contribution due to stored energy by numerically integrating individual terms in the phase-field free energy over the simulation domain and plotting the integrated terms over time. The stored energy decreases as recrystallization occurs, while the energy due to grain boundaries initially increases as the recrystallized grains grow, then decreases as recrystallization ends and grain growth begins (Fig. S2). At the start of the simulations, the total stored energy in the simulation domain is approximately an order of magnitude greater than the total grain boundary energy for 1.5% plastic work converted to stored energy. For larger fractions of plastic work converted to stored energy, the difference between the initial total stored energy and the total grain boundary energy is greater.

The linearity between inverse assumed plastic work conversion fraction and fitted mobility suggests that the assumption $\gamma K \ll f_{stored}$ is valid at or above 3% plastic work conversion for the presently studied Mg-3Zn-0.1Ca alloy at 20% applied strain. Further, since the numerically integrated stored energy is at least an order of magnitude larger than the numerically integrated energy due to GB curvature at the start of the simulations for all considered conversion fractions, the assumption $\gamma K \ll f_{stored}$ may be valid even for plastic work conversion fractions as low as 0.75%. Therefore, we conclude that the methodology presently applied is sufficient to deduce the product of grain boundary mobility and conversion ratio of plastic work to stored energy. If either the mobility or the dislocation density is measured, the other can be deduced based on the results of this study.

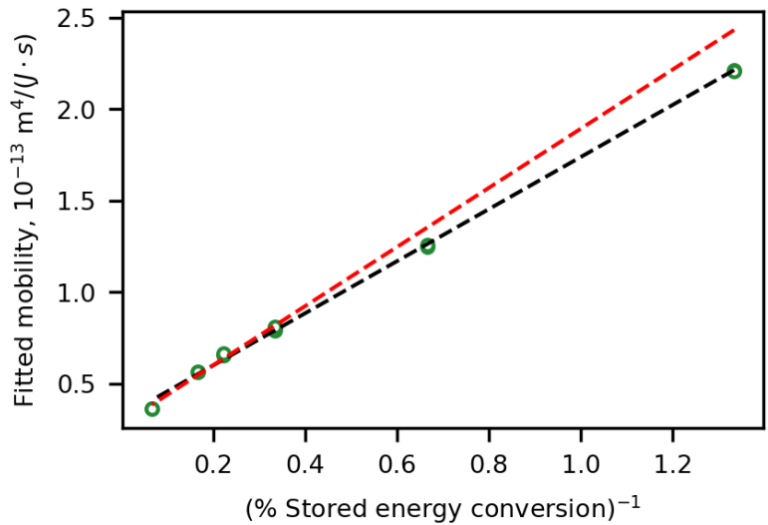

**Fig. S1.** Inverse relationship between fitted mobility and amount of plastic work converted to stored energy, with two linear least-squares fits: one including all the data (black line), and one excluding the data at 0.75% and 1.5% plastic work converted to stored energy (red line). The latter improves the $R^2$ value of the fit, demonstrating the increasingly important contribution of grain boundary surface tension as stored energy decreases.

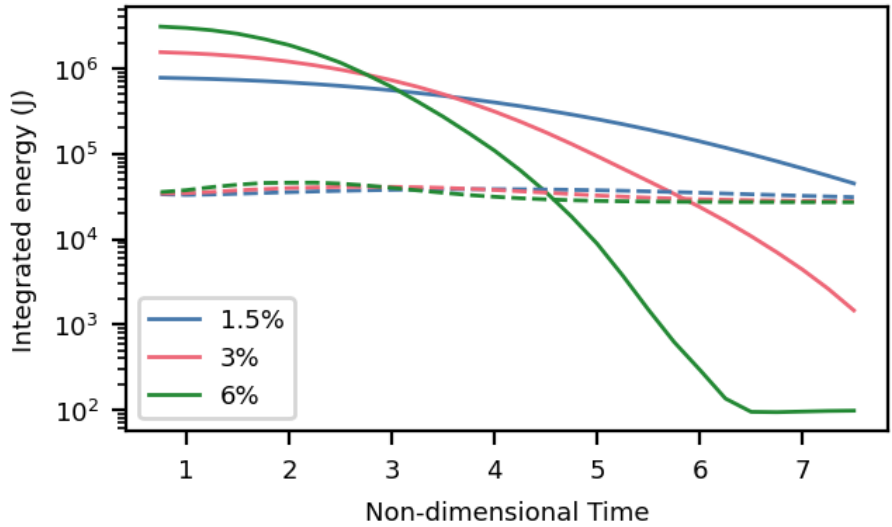

**Fig. S2.** Comparison of total grain boundary energy (dashed curves) and total stored energy (solid curves), integrated over the simulation domain, for different amounts of plastic work converted to stored energy. The simulated conditions are 20% applied strain followed by annealing at 350°C.

*S2. Error in grain boundary energy due to stored energy term*

As discussed by Gentry and Thornton [2], the presence of the stored energy term can impact the grain boundary energy of boundaries between grains with very different stored energies. In the current work, the largest grain-to-grain variation in stored energy is between recrystallized grains (assumed to have zero stored energy) and the most-deformed grains. We examine the order parameter profiles and grain boundary energy for flat pseudo-one-dimensional grain boundaries in bicrystals where one grain has zero stored energy, and the other grain has either zero or constant stored energy. These simulations use the same parameters as in the main article, except the domain size, dimensionless grid spacing, and dimensionless timestep, which are 60×3×3 µm, 0.125, and 0.001, respectively. We examine three different cases in which the rightmost grain is assigned a stored energy of zero while the leftmost grain is assigned a stored energy of 0 MPa, 2 MPa or 4 MPa. This latter case is representative of a "worse case" scenario, considering that largest average stored energy of the simulations in the main article is 4.11 MPa.

The order parameter profiles for the case with zero stored energy in both grains reproduce the expected equilibrium profiles described by a hyperbolic tangent function (Fig. S3a). For the cases with 2 MPa and 4 MPa in the rightmost grain, the steady-state profiles are asymmetric, with the order parameter corresponding to the grain with zero stored energy extending farther, as previously reported by Gentry and Thornton [2] (Figs. S3b-c). The error in grain boundary energy increases as the disparity in stored energy increases (Table S1). The case with a stored energy difference of 4 MPa across the boundary has an error of 24% in the grain boundary energy.

**Table S1.** Error in grain boundary energy due to stored energy difference between grains

| Case | $\gamma_{gb}$ (J / m$^2$) | $\gamma_{gb,calc}$ (J / m$^2$) | Error |
|---|---|---|---|
| Both grains zero stored energy | 0.5 | 0.5 | 0% |
| Zero and 2 MPa stored energy | 0.5 | 0.55 | 10% |
| Zero and 4 MPa stored energy | 0.5 | 0.62 | 24% |

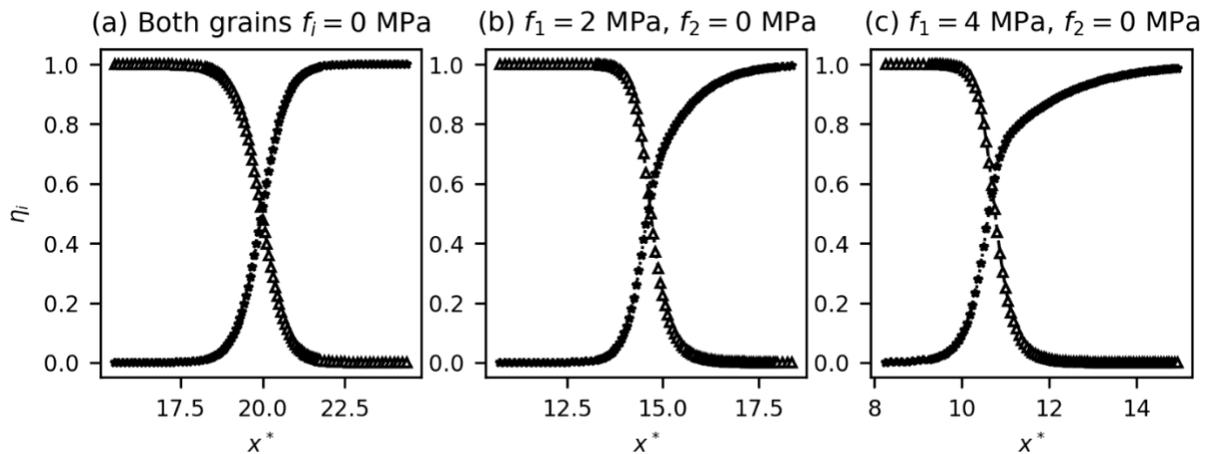

**Fig. S3.** Steady-state order parameter profiles for flat grain boundaries in pseudo-one-dimensional bicrystals, where the left grain ($\eta_1$) has varying stored energy and the right grain ($\eta_2$) has zero stored energy. The distortion of the order parameter profile in the grain with zero stored energy results in an increased grain boundary energy as stored energy disparity across the boundary increases.

*S3. Stereological analysis for approximating grain number density*

In Table 4 of the main article, we report the number density of grains per unit volume observed from the experiments. The micrographs corresponding to each row in Table 4 of the main article are given in Fig. S4. The number densities were determined from stereological analysis, using the method of Saltykov [6-8] with 20 linear size classes. The 2D grain size distribution was obtained from analysis of the EBSD data using the open-source MTEX library [9] in MATLAB as follows. First, the EBSD data in .osc format was loaded, and the grain boundaries identified between grains with a 5° misorientation angle or higher. Grains with a size less than 5 pixels were removed, then the grain boundaries were smoothed. Boundaries within 2° of the exact twinning misorientation were removed, effectively merging the twins into the parent grains. Finally, grains intersecting the image boundary were removed. The areas of the remaining grains were calculated and used to compute the effective radius, defined as the radius of a circle with the same area as the grain.

The distribution of effective radii was used to compute the 2D grain size distribution as a histogram with 20 equal sized bins with size $\Delta R$. The Saltykov method relies on the probability that a given 2D cross section of a sphere belongs to a 3D sphere of a given radius. A 3D sphere of radius $R_i$ can produce cross sections with radii from 0 to $R_i$. The probability, $P_{ij}$, that a cross section with radius in the $j$-th bin belongs to a sphere of 3D radius $R_i$ is

$$P_{ij} = \begin{cases} \frac{1}{R_i}\left(\sqrt{R_i^2 - R_{j-1}^2} - \sqrt{R_i^2 - R_j^2}\right), & R_j \leq R_i \\ 0, & \text{otherwise} \end{cases} \quad (S2)$$

where $R_{j-1}$ and $R_j$ are the lower and upper bounds of the $j$-th size class (bin). Assuming spherical grains, the number density of grains in 2D and 3D can be related by

$$N_A(j) = \sum_{i \geq j}^{N} 2R_i P_{ij} N_v(i) \quad (S3)$$

This can be expressed as a matrix equation of the form $\boldsymbol{N_A} = \boldsymbol{k N_v}$ where $k_{ij} = 2R_i P_{ij}$ is an upper triangular matrix. Solving for vector $\boldsymbol{N_v}$ in equation S3 by inverting $\boldsymbol{k}$ gives the number density per unit volume attributed to each size class. To determine the total number density per unit volume of grains, we simply sum $N_v(i)$ over the index $i$.

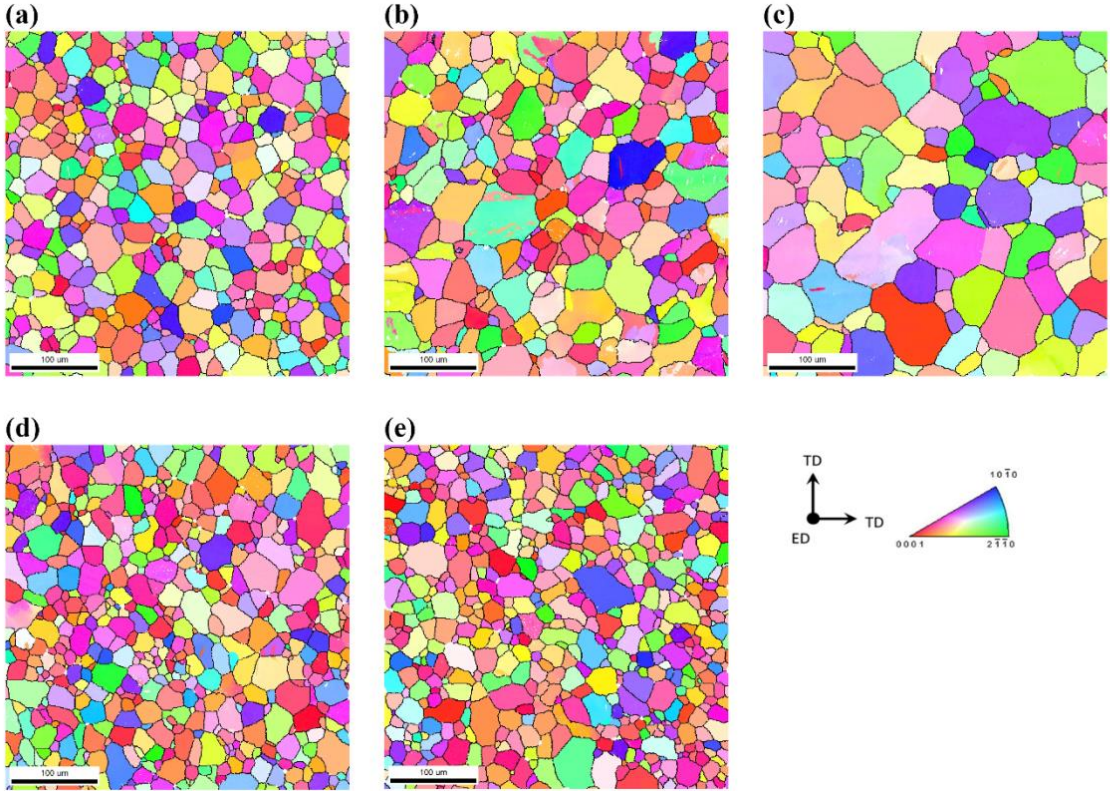

**Fig. S4.** EBSD micrographs of Mg-3Zn-0.1Ca specimens deformed at 200 °C and annealed, corresponding to grain number densities reported in Table 4 of the main article. (a) 20% applied true strain, 350 °C annealing for 150 s. (b) 10% applied true strain, 350 °C annealing for 90 s. (c) 5% applied true strain, 350 °C annealing for 600 s. (d) 20% applied true strain, 310 °C annealing for 600 s. (e) 20% applied true strain, 275 °C annealing for 21600 s.